\documentstyle[aps]{revtex}
\begin{document}
\tightenlines
\draft
\title{Warm inflation with back - reaction: a stochastic approach}
\author{Mauricio Bellini\footnote{E-mail address: mbellini@mdp.edu.ar}}
\address{Departamento de F\'{\i}sica, Facultad de Ciencias Exactas  y
Naturales \\ Universidad Nacional de Mar del Plata, \\
Funes 3350, (7600) Mar del Plata, Buenos Aires, Argentina.}
\maketitle
\begin{abstract}
I study a stochastic approach for warm inflation considering
back - reaction of the metric with the fluctuations of matter field.
This formalism takes
into account the local inhomogeneities of the spacetime in a globally flat
Friedmann - Robertson - Walker metric. The stochastic equations for the
fluctuations of the matter field and metric are obtained. Finally,
the dynamics for the amplitude of these fluctuations in a power - law
expansion for the universe are examined.
\end{abstract}
\vskip 2cm
\noindent
PACS number(s): 98.80.Cq, 04.62.+v \\
\vskip 2cm
Recently, Berera and Fang\cite{3a} showed how thermal fluctuations
may play the dominant role in producing the initial perturbations
during inflation. They invoked slow - roll conditions. This
ingenious idea was extended in some papers\cite{4a} into the warm
inflation scenario. This scenario served as an explicit demonstration that
inflation can occur in presence of a thermal component. However, the
radiation energy density $\rho_r$ must be small with respect to the
matter energy density $\rho_{\varphi}$. More exactly, the kinetic
component of the energy density ($\rho_{kin}$)
must be small with respect to the vacuum
energy density.

An alternative formalism for warm inflation was developed
in previous works\cite{2}. However, in these works is not
taken into account the back - reaction of the metric with the
fluctuations of the matter field. The aim of this letter is to
include the back - reaction of the metric in the formalism previously
developed in\cite{2}.
In the warm inflation scenario the kinetic
energy density ($\rho_{kin}$)
is small with respect to the vacuum energy,
which is given by the density of potential energy $V(\varphi)$
\begin{displaymath}
\rho(\varphi) \sim \rho_m \sim V(\varphi) \gg \rho_{kin}.
\end{displaymath}
where $\rho_{kin} = \rho_r(\varphi)+ {1\over 2} \dot\varphi^2$,
$\rho_m$ is the matter energy density
and
\begin{equation}\label{nue}
\rho_r(\varphi) = \frac{\tau(\varphi)}{ 8H(\varphi)} \dot\varphi^2,
\end{equation}
is the radiation energy density. Here, $\tau(\varphi)$ and $H(\varphi)$
are the $\varphi$ - dependent friction and Hubble parameters.

The density Lagrangian
that describes the warm inflation scenario is
\begin{equation}\label{lag}
{\cal L}(\varphi,\varphi_{,\mu})
= - \sqrt{-g}\left[\frac{R}{16 \pi}
+\frac{1}{2} g^{\mu \nu} \varphi_{,\mu} \varphi_{,\nu}
+V(\varphi) \right]+{\cal L}_{int}.
\end{equation}
Here, $R$ is the scalar curvature, $g^{\mu\nu}$ is the metric
tensor and $g$ is the metric. The Lagrangian ${\cal L}_{int}$
takes into account
the interaction of the field $\varphi$ with other particles of the
thermal bath. The interaction of $\varphi$ with the thermal bath
is represented as a friction parameter in eq. (\ref{nue}).
All particlelike matter which existed before
inflation would have been dispersed by inflation.
If the thermal bath is sufficiently large, it will
act as a heat reservoir which induces fluctuations on the
inflaton field. In warm inflation the mean temperature
of this bath ($T_{ra}$), must be smaller
than the Grand Unified Theories (GUT) one: $T_{ra} <
T_{GUT} \sim 10^{15}$ GeV. This condition implies that
magnetic monopole suppression works effectively.

As in previous works\cite{4,7}, I consider a semiclassical
expansion for the scalar field $\varphi(\vec x,t)$:
$ \varphi(\vec x,t)=\phi_c(t)+\phi(\vec x,t)$.
Here,
$<E|\varphi|E>=\phi_c(t)$ is the expectation value of the operator
$\varphi$ in an arbitrary state $|E>$.
Furthermore, one requires that
$<E|\phi(\vec x,t)|E>$ $=$ $<E|\dot\phi(\vec x,t)|E>=0$.
To consider the quantum fluctuations of the metric, one
takes into account a
quantum perturbed flat Friedmann - Robertson - Walker
(FRW) metric $ ds^2 = - dt^2 + A^2(\vec x,t) d\vec x^2 $
where $A(\vec x,t)$ is a perturbed scale factor of the universe:
$ A(\vec x,t )= a_o \  e^{\int H[\varphi(\vec x,t)] \  dt}$.

Assuming a first order
semiclassical expansion, the Hubble parameter becomes
$H(\varphi) = H_c(\phi_c)
+ \left.{d H(\varphi) \over d \varphi}\right|_{\phi_c} \  \phi
\equiv H_c + H' \phi $. With
this expansion for $H(\varphi)$ one obtains the
expression for the quantum fluctuations of the metric
$h(\vec x,t) \simeq 2 \int^{t} dt' H' \  \phi(\vec x, t')$ (for small
values of $h$), for the metric
$ds^2 = -dt^2 + a^2(t)\left[1+h(\vec x,t)\right] d\vec x^2$.
Due to $<E|h(\vec x,t)|E> = 0$, the
expectation value
for the metric $<E|ds^2|E> = -dt^2 + a^2(t) d\vec x^2 $ gives
a globally flat FRW metric.

The quantum equation of motion for the operator $\varphi$ is
\begin{equation}\label{as}
\ddot\varphi - \frac{1}{<E|A|E>^2}
\nabla^2 \varphi + \left[3 H(\varphi)+
\tau(\varphi)\right] \  \dot\varphi
+ V'(\varphi)=0,
\end{equation}
where $<E|A(\vec x,t)|E>^2 = a^2(t)$ for $A^2(\vec x,t) \simeq
a^2(t) \left[1+h(\vec x,t)\right]$.
The semiclassical Friedmann equation for a globally flat FRW metric is
\begin{equation}\label{as1}
\left<E|H^2(\varphi)|E\right>
=\frac{8 \pi}{3 M^2_p} \left<E| \rho_m(\varphi)+\rho_r(\varphi)|E \right>,
\end{equation}
where the matter and radiation energy densities are
\begin{eqnarray}
\rho_m(\varphi) &=& \frac{\dot\varphi^2}{2} + \frac{1}{2 a^2}
\left(\vec\nabla \varphi \right)^2+ V(\varphi), \\
\rho_r(\varphi) &=& \frac{\tau(\varphi)}{8 H(\varphi)} \dot\varphi^2.
\end{eqnarray}
The $\varphi$ dependent parameter $\tau(\varphi)$ describes
the spatially inhomogeneous friction during the expansion.
This friction becomes from the interaction of the matter field with
the fields of the thermal bath.
The following procedure consist in equating the eqs. (\ref{as}) and
(\ref{as1}) at same order in $\phi$ to obtain the dynamics of the system,
once one makes the semiclassical expansions:
\begin{eqnarray}
&&\varphi = \phi_c + \phi, \\
&& H(\varphi) = H_c(\phi_c) + H' \phi, \\
&& \tau(\varphi) = \tau_c(\phi_c)+\tau' \phi, \\
&& V(\varphi) = V(\phi_c) + V' \phi + 1/2 V'' \phi^2, \\
&& V'(\varphi)= V'(\phi_c)+ V'' \phi + 1/2 V''' \phi^2.
\end{eqnarray}
We consider the eq. (\ref{as}) at zero, first and second order in $\phi$.
These equations with $\tau(\varphi) = \gamma H(\varphi)$
[i.e., for $\tau_c(\phi_c)= \gamma H_c(\phi_c)$ and $\tau'(\phi_c) \phi =
\gamma H'(\phi_c) \phi$] are
\begin{eqnarray}
&& \ddot\phi_c+ 3H_c(\phi_c)[1+\gamma/3] \dot\phi_c
+ V'(\phi_c)=0,\label{bu}\\
&& \ddot\phi - \frac{1}{a^2} \nabla^2 \phi
+ 3 H_c\left[ 1 + \gamma/3 \right]\dot\phi
+  \left[3 H'\left( 1 + \gamma/3\right) \dot\phi_c + V'' \right] \phi =0,  \label{eq}\\
&& 3 H' \left(1 + \gamma/3 \right) \phi \dot\phi + V''' \phi^2 =0, \label{ju}
\end{eqnarray}
where $\gamma $ is a dimensionless constant which is a parameter
of the theory. For $\gamma =0$ one recovers the standard inflation scenario
where the bath is with zero temperature\cite{7}. In this case the Lagrangian
${\cal L}_{int}$ in (\ref{lag}) can be neglected.
The particular choice $\tau_c = \gamma H_c$ for the friction
parameter, becomes from the requirement that the interaction of the
matter field with the particles of the bath must decrease with time for
the thermodynamic equilibrium holds at the end of inflation.
Since $\dot H_c = - {M^2_p \over 4\pi} (H'_c)^2 \left(1+ {\tau_c \over
3 H_c} \right)^{-1} < 0$, the thermodynamic equilibrium is garantized
with the choice $\tau_c = \gamma H_c$. A more general study for the
back - reaction of the metric with the matter field fluctuations
will be developed in a further work\cite{6}.

From eq. (\ref{as1}),
the Friedmann equations at zero, first and second order in $\phi$, are
\begin{eqnarray}
H^2_c(\phi_c)&=&\frac{4\pi}{3 M^2_p}\left[ \left(1+
\frac{\gamma}{4}\right) \dot\phi^2_c+ 2 V(\phi_c)\right], \label{a11}
\\
\left< H_c H' \phi\right> &=& \frac{4 \pi}{3 M^2_p} \left<
\dot\phi_c \dot\phi \left(1+ \gamma/4\right) + V' \phi\right>=0,
\\
\frac{{\cal K}}{a^2}& = &\left<(H')^2 \phi^2 \right> =
\frac{4 \pi}{3 M^2_p} \left< \dot\phi^2 \left(1+ \gamma/4\right)
 +  \frac{1}{a^2} \left(\vec\nabla \phi \right)^2 +
V'' \phi^2 \right>.\label{ja}
\end{eqnarray}
The eq. (\ref{ja}) gives the effective curvature (${\cal K}$) of spacetime
due to the back - reaction of the metric with the fluctuations of the
matter field. Note that this curvature depends on the constant $\gamma $
(i.e., depends on the friction parameter $\tau_c$).
The classical potential is
\begin{equation}
V(\phi_c) = \frac{3 M^2_p}{8\pi}\left[ H^2_c(\phi_c)
 - \frac{M^2_p}{12\pi}\left(H'_c\right)^2 \left(1+\frac{\gamma}{4}
\right)
\left(1+\frac{\gamma}{3}\right)^{-2}\right].
\end{equation}
The eq. (\ref{eq}), with $\dot\phi_c = - \frac{M^2_p}{4\pi} H'_c
\left(1+ \frac{\gamma}{3}\right)^{-1}$ and (\ref{ju}), is
\begin{equation}\label{u}
\ddot \phi - \frac{1}{a^2} \nabla^2 \phi + R(t) \dot\phi =0,
\end{equation}
where
\begin{equation}\label{nue2}
R(t) = 3 H_c \left(1+\frac{\gamma}{3}\right) +
\frac{(1+\gamma/4)}{(1+\gamma/3)} \frac{M^2_p H'_c V''}{(4\pi V'-
3 M^2_p H'_cH_c)}- \frac{3 M^4_p (H')^3}{(4\pi V'-
3 M^2_p H'_cH_c)},
\end{equation}
and $H'_c \equiv {d \over d\phi_c}H_c(\phi_c)$.
One can redefine the quantum fluctuations with the map $\chi =
e^{1/2\int R(t) dt} \phi$, and the eq. (\ref{u}) becomes
\begin{displaymath}
\ddot\chi - \frac{1}{a^2} \nabla^2 \chi - \frac{k^2_o}{a^2} \chi =0,
\end{displaymath}
with
\begin{equation}\label{18'}
k^2_o(t) = a^2 \left[ \frac{1}{2} \left( \frac{R^2}{2}
+ \dot R(t)\right)
\right].
\end{equation}
The redefined quantum fluctuations for the matter field
and the fluctuations of the metric, written as a Fourier expansion, are
\begin{eqnarray}
\chi(\vec x,t) & = & \frac{1}{(2\pi)^{3/2}} \int d^3k \left[
a_k \chi_k + a^{\dagger}_k \chi^*_k \right],\\
h(\vec x,t) & = & \frac{1}{(2\pi)^{3/2}} \int d^3k \left[
a_k h_k + a^{\dagger} h^*_k\right],
\end{eqnarray}
where $\chi_k (\vec x,t)= \xi_k(t)e^{i\vec k.\vec x}$,
$h_k(\vec x,t) ={\mathaccent "707E\xi}_k(t) e^{i\vec k.\vec x}$ and
${\mathaccent "707E\xi}_k(t) = 2 {\large\int} H'(t) \xi_k e^{-1/2
\int^{t} R(t') dt'} \  dt $.
Furthermore $a_k$ and $a^{\dagger}_k$ are
the annihilation and creation operators with commutation
relations $\left[a_k,a^{\dagger}_{k'}\right] = \delta^{(3)}(k-k')$.
The commutation relation between $\chi$ and $\dot\chi$
is $ \left[ \chi(\vec x,t), \dot\chi(\vec x',t) \right] = {\rm i}
\delta^{(3)}(\vec x - \vec x') $, and thus
one obtains $ \left[h(\vec x,t), \dot h(\vec x',t)
\right] = \frac{1}{(2\pi)^3} {\large\int }
d^3k \left( {\mathaccent "707E\xi}_k(t)
\dot{\mathaccent "707E\xi}_k(t)- \dot {\mathaccent "707E\xi}_k(t)
{\mathaccent "707E\xi}_k(t)\right) e^{- i \vec k.(\vec x - \vec x')}$.
The equation of motion for the time dependent modes is
$ \ddot \xi_k(t) + a^{-2}(t) \left[k^2 - k^2_o(t)\right] \xi_k(t)=0 $.

To study the matter and metric fluctuations we can separate
the spectrum in both, the ultraviolet (UV) ($k > \epsilon k_o$)
and the infrared (IR)
($k < \epsilon  k_o$) sectors (where $\epsilon \ll 1$ is a
dimensionless constant). In the IR sector both, $\chi(\vec x,t)$ and
$h(\vec x,t)$ are classical, while in the UV sector
they are quantized. Since
$\ddot \xi_k - \frac{k^2_o}{a^2} \xi_k \simeq 0$,
in the IR sector the
classical stochastic equation that describes the dynamics of the
fluctuations $\chi_{cg}$ is (for $\chi = \chi_{cg}+\chi_S$)
\begin{equation}\label{equa}
\ddot \chi_{cg} - \frac{k^2_o}{a^2} \chi_{cg} = \epsilon \left[
\frac{d}{dt} \left(\dot k_o \eta \right) + 2 \dot k_o \kappa\right],
\end{equation}
where
\begin{eqnarray}
\chi_{cg}(\vec x,t) & = & \frac{1}{(2\pi)^{3/2}}
\int d^3k \  \theta(\epsilon k_o - k)
\left[ a_k \chi_k + a^{\dagger}_k \chi^*_k \right], \\
\eta(\vec x,t) &=& \frac{1}{(2\pi)^{3/2}}
\int d^3 k \  \delta(\epsilon k_o -k)
\left[ a_k \chi_k + a^{\dagger}_k \chi^*_k \right], \\
\kappa(\vec x,t) & = &
\frac{1}{(2\pi)^{3/2}} \int d^3 k \  \delta(\epsilon k_o -k)
\left[ a_k \dot\chi_k + a^{\dagger}_k \dot\chi^*_k \right].
\end{eqnarray}
Note that $k_o$ in eq. (\ref{equa}) depends on the friction parameter
$\tau_c = \gamma H_c$ [see also eqs. (\ref{18'}) and (\ref{nue2})].
When
$(\dot k_o)^2 \left<(\kappa)^2\right> \ll (\ddot k_o)^2 \left<
(\eta)^2 \right>$ (i.e., for $\left|{\dot k_o \dot \xi_{\epsilon k_o} \over
\ddot k_o \xi_{\epsilon k_o}} \right| \ll 1$\cite{7}), we can neglect
the noise $\kappa$ with respect to the another
noise ($\eta$) in (\ref{equa})
and the stochastic equation of motion for the
coarse - grained field that describes the fluctuations of the
metric in the IR sector is (for $h = h_{cg}+h_S$)
\begin{equation}
\ddot h_{cg} = \left[2 \frac{d}{dt}(H')
- H' R(t) \right] e^{-1/2\int R(t)dt}
\chi_{cg}
+ 2 H' e^{-1/2\int R(t) dt} \left[u + \epsilon \dot k_o \eta\right],
\end{equation}
where $\dot u = {k^2_o \over a^2} \chi_{cg}$ and
\begin{displaymath}
h_{cg}(\vec x,t) = \frac{1}{(2\pi)^{3/2}}
\int d^3k \  \theta(\epsilon k_o - k)
\left[ a_k h_k + a^{\dagger}_k h^*_k \right].
\end{displaymath}

On the other hand, the quantum stochastic equation that determines the
dynamics of the quantum fluctuations of the matter field
$\chi_S$ in the UV sector [where it is valid the
equation for the time dependent modes
$\ddot\xi_k + a^{-2} [k^2 -k^2_o(t)] \xi_k=0$ and the commutation
relation is $[\chi_S, \dot\chi_S]={\rm i}\delta^{(3)}(\vec x - \vec x')$],
is
\begin{equation}\label{29}
\ddot \chi_{S} - \frac{k^2_o}{a^2} \chi_S = - \epsilon \left[
\frac{d}{dt} \left(\dot k_o \eta_q \right) + 2 \dot k_o \kappa_q\right],
\end{equation}
where
\begin{eqnarray}
\chi_{S}(\vec x,t) & = & \frac{1}{(2\pi)^{3/2}}
\int d^3k \  \theta(k - \epsilon k_o)
\left[ a_k \chi_k + a^{\dagger}_k \chi^*_k \right], \\
\eta_q(\vec x,t)
& = & \frac{1}{(2\pi)^{3/2}} \int d^3 k \  \delta(k- \epsilon k_o)
\left[ a_k \chi_k + a^{\dagger}_k \chi^*_k \right], \\
\kappa_q(\vec x,t)
& = & \frac{1}{(2\pi)^{3/2}} \int d^3 k \  \delta(k- \epsilon k_o)
\left[ a_k \dot\chi_k + a^{\dagger}_k \dot\chi^*_k \right].
\end{eqnarray}

Now we consider the special case of a power - law expansion of the
universe for which $a \sim (t/t_o)^p$ and $H_c(t) = p/t$. The
squared wavenumber
that separates both, the UV and IR sectors, is
\begin{equation}
k^2_o(t) = H^{-2}_o \  t^{2(p-1)} K^2(\gamma,p),
\end{equation}
where
\begin{equation}
K^2(\gamma,p)= 9/4p^2(1+\gamma/3)^2-
3/2 p(1+\gamma/3)
+\frac{3p^2M^2_p}{2\pi m^2}
\left(
1-\frac{M^2_p}{ 48 \pi^2 m^2}(1+\gamma/4)(1+\gamma/3)^{-2}\right).
\end{equation}
The general solution for the time dependent modes $\xi_k(t)$ is
\begin{equation}
\xi_k(t)  =  A_1 \sqrt{t/t_o} \  H^{(1)}_{\nu}  \left[
\frac{H_o k \  (t/t_o)^{1-p}}{p-1}\right]
+ A_2 \sqrt{t/t_o} \  H^{(2)}_{\nu}
\left[\frac{H_o k \  (t/t_o)^{1-p}}{p-1}\right],
\end{equation}
where $H^{(1)}_{\nu}$ and $H^{(2)}_{\nu}$ are the first and second
species Hankel functions and
$\nu = {1 \over 2(p-1)} \sqrt{1+4 K^2(\gamma,p)}$.
We choose $A_1=0$. For $p > 1$ (i.e., for $\nu > 3/2 $,
$\gamma \ll 1$ and $t\gg 1$), one
obtains the asymptotic solution
\begin{equation}
\xi_k(t)  =  A_2 \sqrt{t/t_o} \left[\frac{1}{\sqrt{2\pi \nu}} e^{\nu}
\left(\frac{H_o k \  (t/t_o)^{1-p}}{2 \nu (p-1)}\right)^{\nu}
 +  {\rm i}  \sqrt{\frac{2}{\pi \nu}} e^{-\nu}
\left( \frac{H_o k \  (t/t_o)^{1-p}}{2 \nu (p-1)}\right)^{-\nu}\right],
\end{equation}
where $e$ denotes the exponential number.
From the commutation condition
$[\chi(\vec x,t), \dot\chi(\vec x',t)]
= {1\over (2\pi)^3} \large{\int} d^3 k \  (\xi_k \dot\xi^*_k
-\dot\xi_k \xi^*_k) \  e^{-i \vec k. (\vec x-\vec x')} =
{\rm i} \delta^{(3)}(\vec x-\vec x')$ one obtains the value
$A_2 = {{\rm i}\over 2} \sqrt{{\pi \over p-1}}$.

In the IR sector
($k \ll {2\nu(p-1)(t/t_o)^{p-1}\over H_o}$), the
asymptotic time dependent modes become
\begin{equation}
\left.\xi_k\right|_{IR} \simeq -\sqrt{t/t_o} \frac{e^{-\nu}}{2}
\sqrt{\frac{2}{\nu (p-1)}}
\left(\frac{H_o k \  (t/t_o)^{1-p}}{2\nu (p-1)}\right)^{-\nu},
\end{equation}
which are asymptotically real.
On the other hand, in the UV sector
the modes are
\begin{equation}
\left.\xi_k\right|_{UV}
\simeq  \sqrt{t/t_o} \left[{\rm i} \frac{e^{\nu}}{2}
\sqrt{\frac{1}{2\nu (p-1)}}
\left(\frac{H_o k (t/t_o)^{1-p}}{2\nu (p-1)}\right)^{\nu}
- \frac{e^{-\nu}}{2} \sqrt{\frac{2}{\nu (p-1)}}
\left(\frac{H_o k (t/t_o)^{1-p}}{2\nu (p-1)}\right)^{-\nu}\right],
\end{equation}
which are complex functions.
The function that defines the
transformation $\chi = e^{1/2\int R(t) dt} \phi$ is
\begin{equation}
R(t) = p \  t^{-1} \left[ 3 (1+\gamma/3)
+ \frac{(1+\gamma/4)}{(1+\gamma/3)}
\frac{M^2_p m^{-2} C}{8\pi C-3M^2_p}
-\frac{3 M^4_p m^{-2}}{32 \pi^2 C  -  12\pi M^2_p}\right],
\end{equation}
where $C= {3 M^2_p\over 8\pi} \left[1- {M^2_p m^{-2} \over 12\pi}
(1+\gamma/4)(1+\gamma/3)^{-2}\right]$.

The squared amplitude
for the fluctuations of the
metric in the IR sector go as
\begin{displaymath}
\left.<h^2_{cg}>\right|_{IR} \sim t^{\nu (p-1)-2(1+M)+3p},
\end{displaymath}
with
$M =2 p \left[ 3 (1+\gamma/3)+ \frac{(1+\gamma/4)}{(1+\gamma/3)}
\frac{M^2_p m^{-2} C}{8\pi C-3M^2_p}
-\frac{3 M^4_p m^{-2}}{32 \pi^2 C  -  12\pi M^2_p}\right]$.
Note for $M > \nu/2(p-1)+3/2p-1$ the amplitude for the
fluctuations of the metric
in the IR sector decreases with time, but for $M <  \nu/2(p-1)+3/2p-1$
these fluctuations increases.
The squared amplitude for the fluctuations
of the matter field go as $<\phi^{2}_{cg}> \sim <h^2_{cg}>$.

At the end of inflation (for $\phi_c = 0$) one obtains $H_c=H_o$ and
the radiation temperature is
\begin{equation}
\frac{T_{ra}}{M_p} \simeq \left(\frac{15}{64 \pi^3}\right)^{1/4}
\left[ \frac{\gamma H_o}{m \  N(T_{ra})(1+\gamma/3)}\right]^{1/4},
\end{equation}
where $N(T_{ra})$ is the number of relativistic degrees of freedom at
temperature $T_{ra}$ (at the end of inflation). For
$N(T_{ra}) = 10^3$ and $H_o \simeq m$
one obtains the cutoff ${T_{re} \over M_p} < 10^{-5}$ for
$\gamma < 10^{-15}$. In this framework the constant $\gamma$ can
be interpreted as a effective coupling constant due to the
interaction between the inflaton
and the particles of the thermal bath.

Finiteness in the UV sector is achieved
by imposing a cutoff at the horizon scale ($k \sim p/t$).
Here,
the amplitude for the fluctuations of the matter field
decreases for $M > \nu (p-2) + 2$ and $\nu > 3/2$, due to
$\left<h^2_S\right> \sim t^{2\nu (p-2) -2M+4}$.

In this letter I have studied a stochastic approach to
the Warm Inflation scenario considering back - reaction of the
metric due to the fluctuations of the matter field $\phi(\vec x,t)$.
In this theory the classical matter field $<E|\varphi|E> = \phi_c$
lead to the expansion of
the universe, while
the fluctuations of the matter field generate an effective curvature
(${\cal K}$), on the background flat FRW metric
[see eq. (\ref{ja})]. However, global curvature of the
spacetime is zero, due to $<E|h(\vec x,t)|E>=0$.
In this framework the early universe
is understood as expanding
regions of the universe with locally different rate of expansion
$H(\varphi) \simeq H_c(\phi_c)
+ H'(\phi_c) \phi(\vec x,t)$, but with expectation value $<E|H(\varphi)|E>=
H_c(\phi_c)$.

To summarize, second order stochastic equations
for the fluctuations of the matter field in both, the IR and UV sectors,
were obtained.
Also, a second order stochastic equation for the
fluctuations of the metric in the IR sector was founded.
Furthermore, a new second order quantum stochastic equation for the
matter field fluctuations in the UV sector was founded [see eq. (\ref{29})].
The evolution for the fluctuations of the matter field depends on
the evolution
of the superhorizon and the scale factor, which depends on the friction
parameter that describes the interaction of the matter field with
the fields in the thermal bath. I find that the amplitudes of both,
the matter and metric fluctuations
decreases with time in the IR sector
(for $p>1$ and $\gamma \ll 1$). This implies that the effects on the
now observational scales should be a soft disturbed flat
FRW spacetime. Such fluctuations
should be explained by the quadrupole anisotropy amplitude in the cosmic
background radiation spectrum. However, this topic is beyond the scope
of this letter.


\begin{thebibliography}{99}
\bibitem{3a} Arjun Berera and Li-Zhi Fang,
Phys. Rev. Lett. {\bf 74},
1912 (1995).
\bibitem{4a} A. Berera, Phys. Rev. Lett.
{\bf 75}, 3218 (1995).\\
A. Berera, Phys. Rev. {\bf D54}, 2519 (1996).
\bibitem{2} Arjun Berera, Phys. Rev. {\bf D55}, 3346 (1997).
\bibitem{4} Mauricio Bellini, Phys. Lett. {\bf 428}, 31 (1998);\\
Nuo. Cim. {\bf B113}, 1481 (1998);\\
Phys. Rev. {\bf D58}, 103518 (1998); \\
Class. Quant. Grav. {\bf 16}, 2393 (1999);\\
M. Bellini, {\em Primordial fluctuations of the metric in the
warm inflation scenario with a more realistic coarse - grained
field}, To appear in Nucl. Phys. B. Preprint gr-qc/9908063.
\bibitem{7} M. Bellini, H. Casini, R. Montemayor, P. Sisterna,
Phys. Rev. {\bf D54}, 7172 (1996).
\bibitem{6} M. Bellini, {\em Warm inflation with coupled quantum
fluctuations}, to be submitted for publication.
\end{thebibliography}
\end{document}